\documentstyle[12pt]{article}
\newcommand{\Gzero}{\mbox{$G^{(0)}$}}
\newcommand{\CGC}{\mbox{$C^{\infty}_{c}(G,{\bf C})$}}

\newcommand{\calE}{\mbox{${\cal E}$}}

\newcommand{\Cstar}{\mbox{$C^*$}}

\newcommand{\calA}{\mbox{${\cal A}$}}

\newcommand{\za}{\mbox{${\cal Z}({\cal A})$}}

\newcommand{\Fi}{\mbox{${\Phi_{g}}$}}

\newcommand{\JMP}{{\em J. Math. Phys.\/}}

\begin{document}

\title{TOWARDS NONCOMMUTATIVE \\ QUANTIZATION OF GRAVITY}
\author{Michael Heller \\
Vatican Observatory V-00120 Vatican City Sate
\footnote{Correspondence address:
ul. Powsta\'nc\'ow Warszawy 13/94, 33-110 Tarn\'ow, Poland \\
E-mail address: mheller@wsd.tarnow.pl}
\and
Wies{\l}aw Sasin \\ Institute of Mathematics, \\ Warsaw
University of Technology, \\ Plac Politechniki 1, 00-661 Warsaw,
Poland }
\maketitle
\begin{abstract}
We propose a mathematical struc\-ture, based on a
non\-com\-mu\-ta\-tive geo\-me\-try, which combines essential
aspects of general relativity with those of quantum mechanics,
and leads to correct ``limiting cases'' of both these physical
theories. We (algebraically) quantize a groupoid constructed on
space-time rather than space-time itself. Both space and time
emerge in the transition process to the commutative case. Our
approach clearly suggests that quantum gravitational observables
should be looked for among correlations of distant phenomena
rather than among local effects. A toy model is computed (based
on a finite group) which predicts the value of ``cosmological
constants'' (in the quantum sector) which vanish when going to
the standard space-time physics.
\end{abstract}
\newpage

\section{Introduction}
There are many theoretical indications that at the fundamental
level, i. e. below Planck's scale, the manifold structure of
space-time breaks down (there are so many hints scattered in the
literature that it is difficult to give a list of references;
for a review see \cite{Henry}) and that, in particular, time
loses its ordinary meaning (see for instance
\cite{ConRov,Rovelli1,Rovelli2}). The problem is that
when we give up the manifold structure so many possibilities are
open, none of them being more natural than others, that we are
left only with our subjective preferences.  It seems that the
situation started to change with the discovery of
noncommutative geometry (see \cite{Connes} and works cited
therein). One could claim that the noncommutative
generalization of the usual geometry is ``natural'' in the
following sense. As it is well, known the manifold structure can
be defined in terms of the algebra of smooth functions
$C^{\infty }(M)$ on a set $M$ (this definition is equivalent to
the more standard one in terms of charts and atlas). The algebra
$C^{\infty }(M)$ is of course commutative, and to drop the
commutativity assumption seems to be a natural step to
undertake. Therefore, we take any (associative) algebra and try
to see what does happen if we proceed, as closely as possible,
along the lines established in the usual differential geometry.
It turns out that the change is radical, but surprisingly many
standard methods can be adapted to the new conceptual context.
In this way, many spaces usually regarded as highly pathological
(e. g.  non-Hausdorff, nonmeasurable) can effectively be
investigated with the help of noncommutative methods. Since
general algebras are often too difficult to deal with one looks
for their suitable representation, and it turns out that such a
representation is provided by the algebra of operators in a
Hilbert space. And here the chain of our motivations closes up.
The idea of noncommutativity appeared first in physics as the
noncommutativity of quantum mechanical observables represented
by operators in a Hilbert space.
\par
Another attractive thing, from our point of view, about
noncommutative spaces is their ``global character''. In a
differentiable manifold $M$ the existence of points is
equivalent to the existence of smooth functions which vanish at
these points; algebraically such functions form maximal ideals
in the algebra $C^{\infty }(M)$ of all smooth functions. In
noncommutative algebras, in general, there are no maximal
ideals, and the concept of point is replaced by that of pure
state which is also familiar from quantum mechanics. In spite of
this, a true dynamics can be done on noncommutative spaces, for
instance in terms of derivations of a given noncommutative
algebra. It is then evident that when one uses such spaces to
model physical processes at the fundamental level the usual idea
of space-time is replaced by something drastically different but
still workable. Moreover, since each noncommutative algebra has
a commutative subalgebra, called its center, the possibility is
always open to go, by restricting to this subalgebra, to the
standard commutative geometry.
\par
The above attractive features of noncommutative geometries, and
--- last but not least --- tangible successes in putting into
the noncommutative framework the standard model of fundamental
interactions \cite{ConLot}, have moti\-vated several attempts at
creating a con\-ceptual basis for the noncommutative quantum
theory of gravity (see, for instance,
\cite{Cham}, \cite{Sitarz}, \cite{Hajac}, \cite{ConG},
\cite{ChamCon1}, \cite{ChamCon2}, \cite{MadMour}).
All these attempts explicitly or tacitly assume that it is the
{\em geometry of space-time\/} which should be made
noncommutative (for review see \cite{Fuzzy}). In \cite{HSL} we
have proposed a scheme for a noncommutative quantization of
gravity the main strategy of which consists in starting --- from
the very beginning --- with an abstract noncommutative space
and obtaining from it the usual space-time geometry via the
correspondence with the classical case. To describe our idea,
let us mention that there is the standard method of obtaining a
noncommutative space from a given commutative one(see
\cite[p. 99-102]{Connes}. Roughly
speaking, a given commutative space, for instance a manifold,
should be presented as the quotient of a groupoid $G$ by an
equivalence relation (which can be given as the action of a
group on $G$), and then one should apply the standard method of
constructing a $C^*$-algebra \calA \ on $G$. This $C^*$-algebra
is in general noncommutative and is a basis for our
noncommutative geometry. In the case of space-time $M$, one
should notice that $M$ can be given as the quotient
$M=E/SO(3,1)$ where $E$ is the total space of the fibre bundle
of frames over $M$.  The point is that, by taking the Cartesian
product $E \times SO(3,1)$, we obtain a groupoid, and the above
method can be directly applied. This will be described in
section 2.  With one important proviso. We shall start, right
from the beginning, with the groupoid $G=E\times \Gamma $, where
$E$ is a suitable space and $\Gamma $ a suitable group,
forgetting about space-time $M$, our aim being to obtain
space-time when going from our noncommutative geometry to the
commutative case.
\par
Another important remark. If we assume that $E$ is a smooth
manifold and $\Gamma $ a Lie group, then the noncommutative
space corresponding to the $C^*$-algebra \calA \ will be
strongly Morita equivalent to the smooth manifold $M=E/\Gamma $.
Since in noncommutative geometry, strong Morita equivalence
plays the role of isomorphism (for definition see \cite[p.
140]{Madore} or \cite[p. 40]{Masson}, it would seem that we have
gained nothing with our construction. But this is not so. Strong
Morita equivalence, being a ``noncommutative isomorphism'',
does not know about points and their neighbourhoods, and
consequently the noncommutative space based on the algebra
\calA \ is equivalent to a smooth manifold ``modulo local
properties''; it truly deserves the name of {\em noncommutative
manifold\/}. In this sense, it is a strong generalization of the
usual manifold concept, and can be used in physics to model
nonlocal processes at the fundamental level.
\par
Our generalization can go even further if we give up the
assumption that $E$ is a smooth manifold. For instance, we could
think of it as of a generalized fibre bundle over a space-time
with singularities such that the fibres over ``singular points''
need not be diffeomorphic to the typical fibre. Even such fibres
are admitted which are reduced to the single point. In
\cite{HelSasSing} it has been shown that in such cases the groupoid
$G$ is quite a regular space, and our construction can proceed
essentially with no changes. However, as the result we obtain a
noncommutative space which is no longer strongly Morita
equivalent to a manifold. Although this case seems to be more
mathematically interesting and more promising from the point of
view of physics (we dealt with it in \cite{HSL}), in the present
paper we shall be concerned with the case where all spaces
involved are assumed to be smooth manifolds. Our motivation for
doing so is that in the present paper we want to introduce our
approach to quantizing gravity as simply as possible, and going
beyond the manifold category would provoke many questions which
at the introductory stage would make things more complicated
rather than smoothing them out.
\par
The organization of our material is the following. In section 2
we construct the Hilbert space for our approach to quantum
gravity. Section 3 summarizes those aspects of noncommutative
geometry which are necessary  to formulate a noncommutative
version of general relativity. Its quantization scheme is
presented in section 4, and the transitions to the usual
space-time geometry, on the one hand, and to the standard
quantum mechanics, on the other hand, are discussed in section
5. In section 6, we check the consistency of our scheme by
computing a simple model in which the groupoid $G$ is a
Cartesian product of a 3-dimensional Minkowski space-time and
the group $\Gamma $ is a finite group $D_4$. In section 7, we
argue that observable quantum gravitational phenomena should be
looked for, as strongly suggested by our scheme, among
correlations between distant measurements rather than among
``local phenomena''.  Section 8 summarizes the paper. Some
overlaps with the material presented in \cite{HSL} were
indispensable not only to make the present paper
self-consistent, but also because our aim in preparing it was to
more carefully discuss physical aspects of our approach than it
had been possible in \cite{HSL} in which the mathematical
foundations were the primary objective.
\par

\section{Groupoid of Fundamental Symmetries}
In this Section we construct the Hilbert space for quantum
gravity.  We start from the direct product $G=E\times \Gamma $
where $E$ is an n-dimensional smooth manifold and $\Gamma $ a
Lie group acting on $E$ (to the right). Elements of $\Gamma $
are ``fundamental symmetries'' of our theory. Heuristically, we
could think of $E$ as of the total space of the fibre bundle of
frames over space-time $M$, and of $\Gamma $ as of its
structural group (a connected component of the Lorentz group).
However, we insist upon starting just from a manifold $E$ and a
group $\Gamma $ of ``fundamental symmetries'', our aim being to
deduce space-time $M$ from our model via the correspondence
principle with macroscopic physics. In the present paper we
leave the group $\Gamma $ unspecified. The correct choice of
$\Gamma $ is left for the future development of the proposed
model, and it should be made on physical grounds. Moreover, it
can turn out that our scheme is too narrow to incorporate all
required physics; in such a case the scheme could be enlarged by
substituting for $\Gamma $ a supergroup or a quantum group (and
suitably modifying the model; preliminary analysis shows that it
is possible).
\par
Our next step will be to regard $G$ as a groupoid. Roughly
speaking, groupoid differs from group by the fact that not all
its elements can be composed with each other (composition can be
done only within certain subsets of the groupoid). For the
precise definition of groupoid see, for instance,
\cite{Renault}; here we shall give a less formal description of
$G$ as a groupoid.
\par
Of course, $G$ is a set of pairs $\gamma =(p,q)$ where $q =
pg,\; p,q \in E,\; g \in \Gamma $ (one can also write $\gamma =
(p,g), \; g \in \Gamma$). We can think of such a pair as of an
arrow starting at $p$ and ending at $q$. This arrow can be
interpreted as a fundamental symmetry operation (the name
``fundamental symmetry'' can be attributed, by only slight abuse
of language, to both elements of $\Gamma $ and elements of $G$).
For $G$ to be a groupoid two its subsets should be
distinguished, namely: $G^{(0)}$ -- the subset of all elements
of $G$ of the from $(p, e)$, $p \in E$, where $e$ is a neutral
element of $\Gamma $, i.e.  the subset of all loops (the loop
being an arrow beginning and ending at $p$); and $G^{(2)}$ --
the subset of all these elements of $G$ which can be composed
with each other; viewed as arrows two elements $\gamma_1, \;
\gamma_2 \in G$ can be composed with each other $\gamma =
\gamma_1 \circ \gamma_2$, if the end of $\gamma_2$ coincides
with the beginning of $\gamma_1$. To formally express properties
of the composition one introduces two following mappings: the
{\em source mapping\/} $$ s: G \rightarrow \Gzero $$ defined by
$$   s(p, q) = p, $$ and the {\em range mapping\/} $$ r: G
\rightarrow \Gzero $$ defined by $$   r(p, q) = q. $$ Then, of
course, $$ G^{(2)} = \{(\gamma_1, \gamma_2) \in G \times G:
s(\gamma_1) = r(\gamma_2)\}, $$ and some natural conditions are
satisfied, for instance $$ s(\gamma_1 \circ \gamma_2 ) =
s(\gamma_2) $$ and $$ r(\gamma_1 \circ \gamma_2) = r(\gamma_1)
$$ for every $\gamma_1, \gamma_2 \in G^{(2)}.$ These conditions
can be easily read from the diagram presenting the composition
$\gamma = \gamma_1 \circ \gamma_2$ in the form of arrows (see
\cite{Connes}, pp. 99-100). We should also notice that each
$\gamma \in G$ has the two-sided inverse $\gamma^{-1}$ such that
$\gamma \gamma^{-1} = r(\gamma )$ and $\gamma^{-1} \gamma =
s(\gamma)$ (for short we omit the composition symbol $\circ $).
$G$ has a natural structure of a fibred space with the fibres
$G_p =
\{p\} \times \Gamma ,\; p \in E$.
\par 
The groupoid $G$ with the above structure is also called the
{\em direct product\/} of $E$ and $\Gamma $, and denoted by $G =
E
\triangleleft \Gamma$. Groupoid $G$ is said to be {\em smooth\/} if
$G$ and \Gzero \ carry differentiable structures such that the
mappings $s$ and $r$ are submersions, and the composition
mapping $\circ : G^{(2)} \rightarrow G$ and the natural
inclusion mapping $\iota : \Gzero \rightarrow G$ are smooth. In
our case, $G$ is evidently a smooth groupoid.
\par
Now, our strategy is the following. First, we shall try to
construct, basing on $G$, a noncommutative differential
geometry which would allow us to introduce generalized
(noncommutative) Einstein's field equation. To this end we
define the algebra $\calA = \CGC $ of smooth compactly supported
complex-valued functions on $G$ with the convolution $$
(a*b)(\gamma ) := \int_{G_p}a(\gamma_1 )b(\gamma_2 ), $$ as
multiplication, where $a,b \in \calA $, and $\gamma = \gamma_1
\gamma_2, \; \gamma , \gamma_1, \gamma_2 \in G_p, \, p\in E $.
If $\Gamma $ is an abelian group the convolution is commutative,
if $\Gamma $ is non-abelian group the convolution is
noncommutative giving rise to a noncommutative geometry (which
we shall construct in the next Section).  \calA \ is also an
involutive algebra with involution defined as $a^*(\gamma ) =
\overline{a(\gamma^{-1})}$.
\par
Basing on the geometry determined by the algebra \calA \ we
shall first define generalized Einstein's equation in the
operator form for derivations of \calA , and then, on each fibre
$G_q$, define square integrable functions equipped with the
suitable Hilbert space structure. The direct sum ${\cal H} =
\bigoplus_{q\in E} L^{2}(G_{q})$ will serve us as a state space
of our quantum mechanics. The modulus squared $|\psi |^2$ of the
``wave function'' $\psi \in L^2(G_q)$ is the probability density
of the ``fundamental symmetry'' $\gamma \in G $ to occur.
\par
The crucial point is to make the noncommutative geometry based
on the algebra $\calA = \CGC $ and the Hilbert space ${\cal H} =
\bigoplus_{q\in E} L^2(G_q)$ to collaborate with each other.
This will be achieved in the following way. First, after solving
the generalized Einstein equation, we complete the algebra
\calA \ to a $C^*$-algebra, and then we find a representation of
this algebra in the Hilbert space ${\cal H}$. Now, we can
develop the theory of quantum gravity by following either the
standard formalism of bounded operators on Hilbert space or the
$C^*$- algebra approach. We shall describe all stages of the
above scheme in the following Sections.

\section{Noncommutative Geome\-try of the Grou\-poid}
As it was demonstrated by Koszul {\cite{Koszul}, and later on
extensively used by others, the differential geometry on a
manifold $M$ can be done in terms of the algebra $C^{\infty
}(M)$ of smooth functions on $M$ and the $C^{\infty
}(M)$-modules of smooth sections of smooth vector bundles over
$M$. The main idea of generalizing the standard differential
geometry is to replace the commutative algebra $C^{\infty }(M)$
by any, non necessarily commutative, associative algebra. In
this way, one obtains a vast generalization of the traditional
geometry but, unfortunately, the generalization is not unique:
at several crucial points one can proceed in various directions,
thus obtaining different versions of noncommutative
differential geometry (see \cite{Dubois-Violette}).  Happily
enough, if we choose the derivation based version of
differential geometry on the smooth groupoid $G=E\times \Gamma$,
the generalization is practically unique. The structure of $G$
turns out to be simple enough to exclude unnecessary
complications and at the same time rich enough to guarantee
interesting results.  The derivation based calculus has been
developed in many works (for instance
\cite{DubViol}-\cite{Masson}).  In the rest of this section we
shall follow \cite{SH}.
\par
Derivation of the algebra $\calA $ is defined to be a linear
transformation (endomorphism) $v: \calA \rightarrow \calA $
satisfying the Leibniz rule $$v(ab)=v(a)b+bv(a),$$ $a, b \in
\calA $. The set of all derivations of \calA \ is denoted by
Der\calA . It is a Lie algebra with respect to the bracket
operation $[u,v]=uv-vu,\; u,v \in$Der\calA . In the case of the
algebra $C^{\infty }(M)$, Der$(C^{\infty }(M)$ is a $C^{\infty
}(M)$-module, and it corresponds to all vector fields on $M$. In
the case of a noncommutative algebra \calA, Der\calA \ is not,
in general, an \calA -module but only a \za -module, where \za \
denotes the center of \calA \ (i.e., the set of all elements of
\calA \ which commute with all elements of \calA ). Der\calA \
can be thought of as a noncommutative counterpart of vectors
fields. It should be emphasized that in the framework of
noncommutative geometry, "vector fields" are, in general,
global objects and, consequently, they cannot be said to consist
of vectors.
\par 
The pair $(\calA, V)$, where $V$ is a \za -submodule of Der\calA
, is called {\em differential algebra\/}. In our case $\calA =
\CGC $, and 
as $V$ we choose those derivations of \calA \ which are
naturally adapted to the structure of $G=E\times \Gamma $ (as a
direct product), i.e. all those $v \in $Der\calA \ which can be
presented in the form $v = v_E + v_{\Gamma }$ where $v_E$ is the
"component" of $v$ parallel to $E$, and $v_{\Gamma }$ the
"component" of $v$ parallel to $\Gamma $. More formally, $v \in
$Der\calA \ is said to be {\em parallel to $E$\/} if, for any
$\alpha \in C^{\infty }(\Gamma )$, $v(\alpha \circ pr_{\Gamma
})=0$ where $pr_{\Gamma }$ is the obvious projection. The set of
all derivations of \calA  \ parallel to $E$ is denoted by Der$_E
\calA $. And analogously for derivations {\em parallel to
${\Gamma } $\/}, denoted by Der$_{\Gamma }\calA $. Therefore, $$
V = {\rm Der}_E\calA \bigoplus {\rm Der}_{\Gamma }\calA . $$ To
proceed further, we must introduce a {\em metric\/}, i.e. a \za
-bilinear non-degenerate symmetric mapping $g: V \times V
\rightarrow \calA $. We chose the metric
\begin{equation} \label{R1}
g = pr_E^*g_E + pr_{\Gamma }^*g_{\Gamma }
\end{equation}
where $g_E$ and $g_{\Gamma }$ are metrics on $E$ and $\Gamma $,
respectively. The above choice of both $V$ and $g$ is the
simplest and the most natural one (it is naturally adapted to
the product structure of $G=E \times \Gamma $) but, if
necessary, we could try other choices as well.
\par
Now, we define the mapping $\Phi_g: V \rightarrow V^*$, where
$V^*$, the dual of $V$, is the set of \za -homomorphism from $V$
to
\calA , by 
$$ \Phi_g(u)(v) = g(u,v), $$ $u,v \in V$. The mappings $\Phi_g$
and $\Phi_g^{-1}$ play the role analogous to that of lowering
and raising indices in the standard tensorial calculus. The set
$V^+$ such that $\Phi_g^{-1}(V^+)=V$ is the set of ``invertible
forms''. In our case, all forms are invertible, i. e. $V^+ =
V^*$.
\par
Now, we define the {\em preconnection\/} $\nabla^{*}:V\times
V\rightarrow V^{*}$ with the help of the usual Koszul formula
\begin{eqnarray*}
(\nabla^{*}_uv)(x)=\frac 12[u(g(v,x))+v(g(u,x))-x(g(u,v))\\
+g(x,[u,v])+g(v,[x,u])-g(u,[v,x])],\end{eqnarray*}
for $u,v,x\in
V$, and the {\em linear connection \/} $\nabla :\, V \times V
\rightarrow V$ by
\[\nabla_uv=\Fi^{-1}(\nabla^{*}_uv).\] 
The {\em curvature\/} of this connection is the operator $
R:\,V^3\rightarrow V$ defined by
\[R(u,x)y=\nabla_u\nabla_xy-\nabla_x\nabla_uy-\nabla_{[
u,x]}y.\]
\par
Since $V$ is a {\em free\/} \za -module we can chose a basis in
it and, for any linear operator $T: V \rightarrow V$ define the
{\em trace\/} of $T$ in the usual way, tr$T =
\sum_{i=1}^{k}T^i_i$ and, consequently, for any fixed pair $x,y
\in V$, the family of operators $R_{xy}: V \rightarrow V$ by
$$R_{xy}(u) = R(u,x)y.$$ The {\em Ricci curvature\/} is ${\bf
ric}(x, y) = {\rm tr}R_{xy}$. Finally, by putting ${\bf ric}(x,
y) = g({\bf R}(x),y)$ we obtain the {\em Ricci operator\/} ${\bf
R}: V \rightarrow V$.  ({\bf R} is the adjoint operator of the
\za -bilinear form ${\bf ric}: V \times V \rightarrow \za $).
This allows us to define the {\em generalized Einstein
equation\/} in the operator form
\begin{equation}
{\bf R}-\frac 1{2\alpha}r{\bf I}+\Lambda {\bf I}=\kappa {\bf
T}\label{R2}
\end{equation}
where $\alpha={\rm t}{\rm r}{\bf I},\, r = {\rm tr}{\bf R}$,
$\Lambda$ and $\kappa$ are constants related to the cosmological
constant and Einstein's gravitational constant, respectively,
and {\bf T} a suitably generalized energy-momentum operator.
Since it could be expected that ``at the fundamental level''
there is only ``pure non-commutative geometry'' we shall assume
that ${\bf T} = 0$, but for the sake of generality we shall keep
$\Lambda $ in the equation (if necessary we can always put
$\Lambda = 0$). Therefore, generalized Einstein's equation
assumes the form
\begin{equation}
{\bf G} = 0
\label{R3} \end{equation}
where ${\bf G} := {\bf R} + 2\Lambda {\bf I} $. It can be easily
seen the set that ker${\bf G} := \{v \in V: {\bf G}(v)=0\}$ is a
\za -submodule of $V$. The differential algebra $(\calA, {\rm ker} {\bf
G})$, where $\calA = \CGC $ will be called {\em Einstein
algebra\/} (or {\em Einstein pair\/}). Such an algebra can be
regarded as a solution of the generalized Einstein equation
(strictly speaking only ker{\bf G} is determined by this
equation).
\par
For $\Lambda \neq 0$, eq. (\ref{R3}) assumes the form $$ {\bf
R}(v)= -2\Lambda {\bf I}(v) $$ If the metric $g$ on the ${\cal
Z}({\cal A})$-module $V$ is given a priori, this equation can be
regarded as the eigenvalue equation for the Ricci operator ${\bf
R}$. In the noncommutative framework metric cannot be given
independently of the module of derivations of a given algebra,
and the noncommutative Einstein equation should determine both
the metric and the module of derivations on which the metric is
defined (it is worthwhile to notice that Madore \cite{Fuzzy}
argues that there is essentially unique metric associated with
each noncumulative differential calculus). If we remember that
derivations can be regarded as counterparts of vector fields,
and consequently as responsible for dynamics (``motions'') of
the system, we could draw the following analogy with the
standard case. Just like in the usual general relativity it is
impossible first to specify the distribution and motions of
matter and then from this to compute the structure of space-time
(see, e.g.,
\cite[p. 84]{Stephani}), similarly, in our case, motion (derivations)
and geometry (metric) are so closely dynamically linked with
each other that they can only de determined simultaneously.
\par
In this way, we have obtained the noncommutative version of
general relativity (not yet quantum gravity theory). To solve
eq.  (\ref{R3}) or eq. (\ref{R2}) is a difficult task, but we
can show that it has many solutions. Indeed, let $(M,g)$ be a
solution of the usual Einstein's equation. We construct the
orthonormal frame bundle $\pi : OM \rightarrow M$ over $M$ with
SO(3, 1) as its structural group, and form the groupoid $G = OM
\triangleleft \, $SO(3,1). For the pair $(\CGC, {\rm ker}{\bf
G})$, which is a solution of eq. (\ref{R2}) or (\ref{R3}), there
exists the pair $(C^{\infty }(M), {\rm ker}{\bf \tilde{G}})$
where ${\bf \tilde {G}}$ is the usual Einstein tensor written in
the operator form, i.e. with one index up and one index down,
such that the algebras \CGC \ and $C^{\infty }(M)$ are strongly
Morita equivalent. Strong Morita equivalence plays the role of
isomorphism for noncommutative algebras
\cite{Connes,Madore}. The fact that the algebras \CGC \ and
$C^{\infty }(M)$ are strongly Morita equivalent means that, from
the point of view of the noncommutative algebra, they contain
the same information (let us notice, however, that \CGC \
ignores local properties of $C^{\infty }(M)$). We have shown,
therefore, that for every solution of the usual Einstein
equation there exists the solution of the generalized Einstein
equation such that both solutions are Morita equivalent. Of
course, not all solutions of the generalized Einstein equation
are generated in this way.
\par

\section{Quantization of Noncommutative General Relativity} In
the present Section $(\calA, {\rm ker} {\bf G})$ is an Einstein
algebra. We remind that \calA \ is an involutive algebra with
the involution defined as $a^*(\gamma ) =
\overline{(\gamma^{-1})},\; a\in \calA , \gamma \in G$, and
convolution $(a*b)(\gamma ) = int_{G_p}a(\gamma_1) b(\gamma_2 )$
as multiplication. Now, our aim is to extend \calA \ to a
$C^*$-algebra and quantize it with the help of the standard
algebraic method (see, for instance,
\cite{Thirring}).
\par
By applying the theorem proved by Connes \cite{ConSur} to our
case, we learn that the involutive algebra $\calA = \CGC $, for
each $q \in \Gzero $, has the representation $\pi_q$ in a
Hilbert space ${\cal H} = L^2(G_q)$ $$
\pi_q: \calA \rightarrow {\cal B}({\cal H}),
$$ where ${\cal B}({\cal H})$ denotes the algebra of bounded
operators on ${\cal H}$, given by
\begin{equation} \label{Conrepr}
(\pi_q(a)\psi )(\gamma ) =\int_{G_q}a(\gamma_1)\psi (\gamma_
1^{-1}\gamma ),
\end{equation}
$\gamma = \gamma_1 \circ \gamma_2, \; \gamma , \gamma_1 ,
\gamma_2
\in G_q, \; \psi \in L^2(G_q), \; a \in \cal A $, and that the
completion of \calA \ with respect to the norm $$
\parallel a\parallel\,={\rm sup}_{q\in\Gzero
}\parallel\pi_q(a)\parallel $$ is a $C^*$-algebra. We shall
denote it by ${\cal E}$ and call {\em Einstein $C^*$-algebra\/}.
\par
Now, we can formulate postulates of our noncommutative theory
of quantum gravity.
\par
{\em Postulate 1.\/} A quantum gravitational system is
represented by an Einstein $C^*$-algebra ${\cal E}$, and its
observables by Hermitian elements of ${\cal E}$ (the set of all
Hermitian elements of \calE \ will be denoted by $\calE _H$).
\par
We speak of ``observables'' in the quantum gravity regime by
analogy with the standard quantum mechanics. Whether these
``observables'' leave traces in the macroscopic world remains to
be seen (we come back to this question in Sec. 5). By the same
analogy we can say that the spectrum of a Hermitian element of
\calE \ represents possible measurement results of this observable.
\par
{\em Postulate 2.\/} Let  ${\cal S}$ denote the set of all
states of the algebra \calE ; elements of ${\cal S}$ represent
states of the system and pure states of \calE \ represent pure
states of the system.
\par
{\em Postulate 3.\/} If $a \in \calE_H$ \ and $\phi \in {\cal
S}$ then $\phi (a)$ is the expectation value of the observable
$a$ when the system is in the state $\phi $.
\par
We remind that {\em states\/} of \calE \ are defined to be
positive linear functionals $\phi $ on \calE \ such that
$\parallel \phi
\parallel = 1$. Convex combinations of states are states. A state
which cannot be expressed as a convex combination of other
states is said to be a {\em pure state\/}.
\par
The above three postulates are in the analogy with the standard
\Cstar -algebraic approach to quantum mechanics, the fourth
postulate is a new ingredient of the noncommutative
quantization of gravity.
\par
{\em Postulate 4.\/} The dynamical equation of the system
described by \calE \ is
\begin{equation}
i \hbar \pi_q(v(a)) = [\pi_q(a), F]
\label{dyneq} \end{equation}
for every $q \in \Gzero $. Let us notice that here $v \in {\bf
G}$, and in this way generalized Einstein's equation (\ref{R3})
is coupled to quantum dynamical equation (\ref{dyneq}). $F$ is a
Fredholm operator, i.e. an operator $F: {\cal H} \rightarrow
{\cal H}$ such that $F({\cal H})$ is closed and the dimensions
of its kernel and cokernel are finite. Together with the group
$\Gamma $, the operator $F$ is a ``free entry'' of our
quantization scheme. It should be specified on physical grounds.
To solve eq. (\ref{dyneq}) means to find $a \in {\rm ker}{\bf G}
\subset \calE $ such that $\pi_q(v(a))$ for $v \in 
{\rm Der}\calE $ , would give the same result as $-i/\hbar
[\pi_q(a), F]$ when acting on $\psi \in L^2(G_q)$.
\par
Equation (\ref{dyneq}) is a noncommutative counterpart of the
Schr\"odinger equation in the Heisenberg picture of the usual
quantum mechanics $$ i\hbar(\frac{d}{dt}\hat{A}(t)) =
[\hat{A}(t), H] $$ where $H$ is the Hamilton operator, in which
state vectors are independent of time and all time dependence
goes to operators.  Since in the noncommutative framework the
standard concept of time breaks down, the dynamics of the system
is expressed in terms of derivation of the Einstein algebra.
This remark could be elaborated in the following way. Let {\bf
S} be the set of all elements of the Einstein algebra \calE \
satisfying eq. (\ref{dyneq}), and let us consider the set
$\pi_q({\bf S}) \subset {\cal B}({\cal H})$. Let further
$(\pi_q({\bf S}))^{\prime \prime }$ be the commutant of the
commutant of $\pi_q({\bf S})$ (we remaind that the {\em
commutant\/} of a subset ${\cal M}$ of an algebra \calA \ is
defined to be the set of all elements of \calA \ which commute
with all elements of ${\cal M}$). $(\pi_q({\bf S}))^{\prime
\prime }$ is the smallest von Neumann algebra generated by the
``space of solutions'' $\pi_q({\bf S})$ (see \cite[p.
14]{Masson}). We remind that the subset ${\cal N}$ of ${\cal
B}({\cal H})$ is said to be a {\em von Neumann algebra\/}  if
${\cal N} = {\cal N}^{\prime \prime }$. It can be argued that in
the context of noncommutative geometries von Neumann algebras
encode dynamical aspects of the system in question. Very roughly
speaking, an algebra of operators in a separable Hilbert space
is a von Neumann algebra if it commutes with unitary operators
in this Hilbert space and, as it is well known from the standard
quantum mechanics, unitary operators are responsible for the
dynamical evolution of the system. In the noncommutative
context there is no obvious counterpart of the local time
concept, and it is precisely von Neumann algebra that can be
regarded as implementing the abstract idea of dynamics (see also
\cite{ConRov}).
\par
Equation (\ref{dyneq}) acts on that Hilbert space $L^2(G_q)$.
This space should be regarded as a counterpart of the Hilbert
space in the position representation in quantum mechanics.
However, now the ``position space'' is more abstract: the
quantity $|\psi (\gamma )|^2$ is the probability density of the
``fundamental symmetry'' $\gamma \in G_q$ to occur.
\par
It might turn out that in order to make eq. (\ref{dyneq})
manageable we would have to impose on it some further
conditions, for instance to assume that the triple $(\calA ,
{\cal H}, F)$ is a Fredholm module or to assume some its
summability properties (see \cite[pp.  288-291]{Connes},
\cite[pp. 117-120]{Masson}).
\par
To substantiate our approach we should show that it reproduces,
in a suitable limit, the usual general relativity and the usual
quantum mechanics. We shall demonstrate this in the subsequent
Section.
\par

\section{Transition to General Relati\-vity \newline and
Quan\-tum Mecha\-nics} The ``canonical way'' of obtaining the
commutative geometry from a noncommutative one is to restrict
the corresponding noncommutative algebra \calA \ to its center
\za . In our case this restriction could be interpreted in the
following manner. The noncommutative algebra \calA \ can be
thought of as a ``deformation'' of a commutative algebra of
``classical observables'' with the Planck constant $\hbar $ as a
``deformation parameter'', typically $[a, b] = i\hbar {\bf 1}$,
for two noncommuting elements $a$ and $b$ of \calA
. Since the center \za \ of \calA \ is the set of elements of \cal
A \ which commute with all elements of \calA , the transition
from
\calA \ to \za \  can be implemented by postulating $\hbar
\rightarrow 0$. This means that quantum effects are negligible and
should reduce our theory to the usual theory of general
relativity.  We shall show below that this is indeed the case.
\par
\za \ is of course a commutative algebra with convolution as
multiplication (since \za \ is a subalgebra of \calA ). Let
$\za^{\#}$ be the set of all characters of \za , i.e. the set of
all *-homomorphisms from \za \ to {\bf C}. On the strength of
the Gel'fand theorem the algebra \za \ is isomorphic with the
algebra of continuous functions on the groupoid $G$ (with the
usual multiplication). This algebra is given by the Gel'fand
representation $$  \rho^{\za }: \za \rightarrow {\bf C}^{\za
^{\#}}  $$ defined by $$ \rho ^{\za }(a)(\chi ) = \chi (a)  $$
where $a\in \za , \; \chi \in \za ^{\# }$, and $\za ^{\# }$ can
be identified with $G$. The algebra $\rho^{\za }(\za )$ consists
of continuous functions on $G$, but since $G$ is a smooth
manifold we can assume that these functions are smooth (if
necessary we can restrict this algebra to the subalgebra of
smooth functions) (see
\cite{Palais}). We shall denote the algebra of these functions by
$G^{\infty }$. As it is well known, there is the bijection $$
\za ^{\#} \rightarrow  {\rm Spec}\za ,  $$ where Spec\za \
denotes the set of maximal ideals of \za , given by $$ \chi
\mapsto {\rm ker}\chi , $$ $\chi \in \za^\# $. Each maximal
ideal ker$\chi $ determines a point of $G$  (such a point is
given by the set of functions belonging to $G^{\infty }$
vanishing at this point). We remember that points of $G$ are
``fundamental symmetries'' of our theory. In this way, the full
geometry of the groupoid is given by the pair $(G, G^{\infty
})$.
\par
Since the prototype of our groupoid was the Cartesian product $G
= E \times \Gamma $ (see Introduction) where $E$ was supposed to
be the total space of the frame bundle over space-time $M$, we
recover $M$ by forming, first, the quotient $E  = G/\Gamma $,
and, second, $M = E/\Gamma $ (or $M = (E/\Gamma )/\Gamma $, see
below for the detailed construction). Generalized Einstein's
equation (\ref{R2}) ``projected down'' in this way to space-time
$M$ gives the usual Einstein equation of general relativity.
\par
Now, we shall discuss the transition from our noncommutative
theory to the usual quantum mechanics. To do this, we assume
that the gravitational field is weak so that quantum gravity
effects can be neglected. This means that in dynamical equation
(\ref{dyneq}) the assumption that $v \in $ker{\bf G} can be
omitted, i.e. generalized Einstein's equation is decoupled from
ordinary quantum effects.
\par
Equation (\ref{dyneq}) is defined on the Hilbert space
$\bigoplus_{q\in E} L^2(G_q)$. We want to ``project it down'' to
the more usual Hilbert space $L^2(M)$. We do that, essentially,
as above, by forming the ``double quotient'' $(E/\Gamma)/\Gamma
$. However, more practical way is the following.
\par
Let $(p_1,g_1), \; (p_2, g_2)\in G$, $p_1,\,p_2 \in E,\; g_1,\,
g_2 \in \Gamma $. We define the equivalence relation $$
(p_1,g_1) \sim (p_2,g_2) \Leftrightarrow \exists_{g\in
\Gamma} \, p_2 = p_1g. $$
Then we consider only those ``wave functions'' $\tilde{\psi }$
which have the following invariance property $$  \tilde{\psi
}(p_1, g_1) = \tilde{\psi }(p_2, g_2) $$ ($\tilde{\psi }$ is
constant on equivalence classes of $\sim $).  It can be easily
seen that equation (\ref{dyneq}) restricted to functions
$\tilde{\psi }$ is essentially the Schr\"odinger equation of
quantum mechanics in its Heisenberg picture provided that the
Fredholm operator $F$ is correctly chosen to reproduce the
Hamiltonian of the system. This, of course, had to be expected.
In our noncommutative quantum gravity theory there is no
concepts of points and time instants (space and time are somehow
hidden in the subalgebra of \za ); the standard concept of
space-time appears only in the transition process to the
standard physics. No wonder that we obtain the Heisenberg
picture in which state vectors are time independent and all time
dependance goes to operators.
\par

\section{A Simple Example}
In this Section we shall analyse a simple model of our scheme to
quantize gravity.  The basis of this model is the groupoid
$G=E\times D_4$, where $E$ is the total space of the frame
bundle over a three-dimensional Minkowski space-time $ M^3$ and
$D_4$ is a group consisting of 4 rotations by the angle $
\pi /2$ and 4 reflections 
with respect to two directions crossing each other at the
origin.  If $ r$ denotes rotation and $s$ reflection, the
following relations are assumed to be satisfied
\[r^4=1,\;\;s^2=1,\;\;srs=r^{-1}.\]
$D_4$ is a finite noncommutative subgroup of the group SU(2).  $
D_4$ acts on $E$ (to the right) on the plane $(x,y)$ leaving the
$ t$-axis fixed.
\par
For every vector field $X\in {\cal X}(M^3)$ on the Minkowski
space-time $ M^3$, there exists its lifting to $\bar {X}\in
{\cal X}(G )$ to $G$.  It can be easily seen that the vector
field $\bar {X}$ is constant on the fibres (of $ G$) parallel to
$M^3$.  All such fields form a ${\cal Z}({\cal A})$-submodule $
V_E$ of the ${\cal Z}({\cal A})$-module ${\rm D} {\rm e}{\rm
r}({\cal A})$, where ${\cal A}=C^{\infty}(G)$.  Analogously, we
have a $ {\cal Z}({\cal A})$-submodule $V_{D_4}$ of the ${\cal
Z}({\cal A})$-module ${\rm D}{\rm e}{\rm r} ({\cal A})$.  Our
simple model will be based on the differential algebra $({\cal
A},V)$, where $V=V_E\oplus V_{ D_4},\:V\subset {\rm D}{\rm
e}{\rm r}({\cal A} )$.  Accordingly, the algebra $C^{\infty}(G)$
can be ``decomposed'' into the algebras $ C^{\infty}(E)$ and
{\bf C$[D_4]$,} where $C^{\infty}(E)=\iota_E^{*}(C^{\infty}(G))$
and $ {\bf C}[D_4]=\iota_{D_4}^{*}(C^{\infty}(G))$, $
\iota^{}_E$ and $\iota_{D_4}$ being natural 
embeddings of $E$ and $D_4$ into $G$, respectively.
\par
In agreement with formula (1) (Sec.  3), we assume a metric on
the ${\cal Z}({\cal A})$-module $V$ of the form
\[g=pr_E^{*}\circ\pi_1^{*}\eta +pr_{D_4}^{*}g_{
D_4},\] where $\eta$ and $g_{D_4}$ are the Minkowski metric on $
M^3$ and a metric on the group $D_4,$ respectively; $pr_E$ and
$pr_{D_ 4}$ are obvious projections from the groupoid, and
$\pi_1$ is the canonical projection from $ E$ to $M^3$.
\par
Since the ``parallel geometry'' (geometry based on $ V_E$) is
rather obvious (see below), we shall focus on the ``vertical
geometry'' (geometry of $D_4)$.  As it is well known, the group
algebra ${\bf C} [D_4]$ can be constructed in the following way
\[{\bf C}[D_4]=\{\sum_{i=1}^8c_iA_i:c_i\in {\bf C}
,A_i\in D_4,\,i=1,\ldots ,8\}\] (8 is the rank of $D_4$), with
the usual addition and convolution as multiplication. For any
finite group $\Gamma$ there exists an isomorphism
\[T:\:{\bf C}[\Gamma ]\rightarrow\Pi_{i=1}^kM_{
n_i}({\bf C}),\] where $M_{n_i}({\bf C})$ are $n_i\times n_i$
matrices and $ i$ runs over all irreducible representations of
$\Gamma ,$ such that $T(\varphi *\psi )=T(\varphi )\cdot T(\psi
)$ with asterisk denoting convolution and dot the usual matrix
multiplication.  The group $D_4$ has 4 irreducible
representations of rank 1, and 1 irreducible representation of
rank 2 \cite{Serre}. Therefore, in our case, the above
isomorphism assumes the form
\[T:\:{\bf C}[D_4]\rightarrow {\bf C}\oplus {\bf C}
\oplus {\bf C}\oplus {\bf C}\oplus M_2({\bf C}
)\] given by
\[T=(\lambda_1,\lambda_2,\lambda_3,\lambda_4,
\rho^1),\]
where $\lambda_1,\lambda_2,\lambda_3,\lambda_ 4$ are the rank 1
irreducible representations of $ D_4$, and $\rho^1$ is the rank
2 irreducible representation of $ D_4$.
\par
The set of derivations ${\rm D}{\rm e}{\rm r} ({\bf C}[D_4])$ of
the algebra $ ${\bf C$[D_4 ]$} is isomorphic with ${\rm D}{\rm
e}{\rm r}(M_2({\bf C}))$. It can be shown, by the
straightforward computation, that if $(e_{\alpha}),\,\alpha
=1,\ldots ,n^2 -1$, is a basis of the Lie algebra su(n) of the
Lie group SU(n), and if $c^{\gamma}_{\alpha\beta}$, $
\alpha ,\beta ,\gamma =1,\ldots ,n^2-1,$ are structure constants 
of su(n) with respect to the basis $(e_{\alpha} )$, then
$c_{\alpha\beta}^{\gamma}$ are also structure constants of the
Lie algebra ${\rm D}{\rm e}{\rm r}(M_n({\bf C})$ with respect to
the basis $ ({\rm a}{\rm d}e_{\alpha})$, $\alpha =1,\ldots
,n^2-1.$ In our case, we choose the following basis for su(2)
\[e_1={\rm a}{\rm d}\frac i2\sigma_1,\:e_2={\rm a}
{\rm d}\frac i2\sigma_2,\:e_3={\rm a}{\rm d}\frac i2\sigma_3,\]
where $\sigma_1,\,\sigma_2,\,\sigma_3$ are the usual Pauli
matrices, and the structure constants assume the simple form
$c_{\alpha\beta}^{
\gamma}=\epsilon_{\alpha\beta\gamma}$, i.e. they are equal 1 
for even permutations, and 0 otherwise (see, for instance
\cite[p. 183]{3Panie}).
\par
Now, we chose the metric $g_{D_4}:\,V_{D_4}\times
V_{D_4}\rightarrow {\cal Z}({\bf C}$$[D_4]$) defined by
\[(g_{D_4})_{11}=k{\bf I},\;(g_{D_4})_{ij}=\delta_{
ij}{\bf I},\,{\rm i}{\rm f}\,i\neq 1,\,j\neq 1,\] where $k\in
{\bf R}$, and develop differential geometry as in Sec. 4. As we
shall see below, even such a trivial deviation from the
``Euclidean metric'' gives us interesting insights into the
nature of the problem at hand.
\par
Straightforward computations give the following non-vanishing
components of the Ricci operator \[R^1_1=-\frac k2,\]
\[R^2_2=R^3_3=\frac k2-1.\]
\par
It can be easily seen that{\bf ,} in this case{\bf ,} the
Einstein equation {\bf R$_{ D_4}(w)=0$ } is satisfied only for
$w=0$, $w\in V_{D_4}$. To find non-trivial solutions we should
try the Einstein equation with the cosmological constant. For
$w=w^iv_i$ it can be written in the form
\[({\bf R}_{D_4}+2\Lambda {\rm i}{\rm d})w^iv_
i=0\label{eigen}.\] This is the eigenvalue equation for the
operator $ {\bf R}_{D_4}$. The eigenvalues $\lambda =2\Lambda$
can be easily found. To find them we should distinguish two
cases.
\par
{\sl Case 1:} $k\neq 1$. The eigenvalues are: $
\Lambda_1=\frac k4$ and $\Lambda_2=\frac 12-\frac 
k4$.
\par
The eigenvectors corresponding to the eigenvalue $
\Lambda_1$ are of the form 
$w=tv_i=t{\rm a}{\rm d}\frac i2\sigma_1$, where $ t\in {\bf C}$.
Therefore, for $\Lambda_1$ the solution of the ``vertical part''
of Einstein equation (3) is
\[{\rm k}{\rm e}{\rm r}({\bf G}_{D_4})=\{tv_1
:\,t\in {\bf C}\}\] where {\bf G$_{D_4}={\bf R}_{D_4}+2\Lambda
{\bf I}$.} And correspondingly for the eigenvalue $
\Lambda_2$ one has 
the solution
\[{\rm k}{\rm e}{\rm r}({\bf G}_{D_4})=\{rv_2
+sv_3:\,r,s\in {\bf C}\}.\]
\par
It is a remarkable fact that our simple model requires, for its
consistency, the existence of two ``cosmological constants'' in
the quantum sector of the model (i.e., in its ``vertical
geometry''), and predicts their values. These ``cosmological
constants'' are eigenvalues of the Ricci operator ${\bf
R}_{D_4}$ (up to a constant factor). Of course, because of the
``toy'' character of the model, these values are rather
symbolic.
\par
{\sl Case 2:$ $} $k=1.$ There is only one eigenvalue $
\Lambda =\frac 14$, and correspondingly 
one has \[{\rm k}{\rm e}{\rm r}({\bf G}_{D_4})=\{tv_1 :\,s\in
{\bf C}\}.\]
\par
Now, we must return to the ``parallel geometry''.  We shall
consider the algebra ${\cal A}_E=\pi_2^{*}(C^{\infty}( E))$
where $\pi_2$ is the natural projection from $G=E\times D_4$ to
$E$.  Since in our case the frame bundle $
\pi_1:\,$$E\rightarrow M^3$ is a 
trivial bundle, there is the natural embedding $
\iota_k:\,M\rightarrow E$, $k\in\Gamma$, 
where $\Gamma$ is the structural group of this bundle, i.e.  the
Lorentz group.  There is also another natural embedding $
j_h:\,E\rightarrow G$, $h\in D_4$. With the help of these two
embeddings we ``push forward'' the basis
$(\partial_t,\partial_x,\partial_y)$ in $M^3$ to the basis $
(\bar{\partial}_t,\bar{\partial}_x,\bar{\partial}_ y)$ in $G$.
In this way, we obtain the ${\cal Z}({\cal A})$-module of
derivations
\[V_E=\{\alpha^1\bar{\partial}_t+\alpha^2\bar{
\partial}_x+\alpha^3\bar{\partial}_z:\;\mbox{$
\alpha^1,\alpha^2,\alpha^3\in {\cal A}_E$}\}.\]
We equip this module with the metric $\bar{\eta}$ lifted from
the Minkowski metric $\eta$ on $M^3$, i.e. $
\bar{\eta }=\tau^{*}\eta$ where $\tau =\pi_1\circ
\pi_2$. It 
can be easily seen that in fact $\bar{\eta}$ is also Minkowski
metric. Indeed, let $\bar {X},\bar {Y}\in V_E$, then $\bar{\eta
} (\bar {X},\bar {Y})=(\tau^{*}\eta )(\bar {X},
\bar {Y})=\eta (\tau_{*}\bar {X},\tau_{*}\bar {
Y})=\eta (X,Y)$. Therefore, we have \[{\bf R}_E=0.\]
\par
Now, we can consider the algebra ${\cal A}=C^{
\infty}(E\times D_4)$ together with the 
${\cal Z}({\cal A})$-module
\[V=V_E\oplus V_{D_4}=\]
\[=\{\alpha^a\bar{\partial}_a+\beta^i\bar {v}_
i:\;\alpha^a,\beta^i\in {\cal A},\;a=0,1,2;\, i=1,2,3\}.\]
\par
We should now see how the generalized Einstein equation
interacts with the quantum dynamical equation (5). As an example
let us consider case 1 when, in the ``vertical geometry'', the
metric coefficient $k\neq 1$. Since we postulate that the
derivation $v$ in the left hand side of eq. (5) should be a
solution of generalized Einstein equation, eq. (5) splits into
two equations
\[i\hbar\pi_q((\alpha^a\bar{\partial}_a+t{\rm a}
{\rm d}\frac i2\sigma_1)(a))=[\pi_q(a),F],\] and
\[i\hbar\pi_q((\alpha^a\partial_a+r{\rm a}{\rm d}\frac 
i2\sigma_2+s{\rm a}{\rm d}\frac i2\sigma_3)(a ))=[\pi_q,F].\]
When a Fredholm operator $F$ is given, these equations should be
solved for $a\in {\cal A}$.  \par\ In agreement with the
discussion of Sec.  5, in order to obtain classical case we must
restrict the algebra $ {\cal A}=C^{\infty}(G)$ to the algebra
${\cal A}_{proj}$.  In such a case, the ``vertical geometry''
projects to zero, and we are left with the ordinary Minkowski
space-time $M^3$ and the corresponding Einstein field equation $
{\bf R}=0$ (this effect, in the considered model, is trivial
since the ``parallel geometry'' has been obtained by lifting the
Minkowski geometry to the groupoid $G$).  Let us notice that,
even in this toy model, we have an interesting result:  in the
noncommutative regime the kind of cosmological constants appear
(as eigenvalues of the Ricci operator for the ``quantum
sector'') which vanish if we go to the classical case.
\par
\section{Observables and Their Eigenvalues}
In postulate 2 of our quantization scheme we have identified
quantum gravity observables with the Hermitian elements of the
algebra \calA \ by following strict analogy with the
$C^*$-algebraic quantization of the usual quantum mechanics.
However, from the experimental point of view we are interested
only in those observables which leave some traces in the
macroscopic world and thus have chances to be detected. As we
have seen in the preceding section, such observables must belong
to ${\cal A}_{proj}$. Let $a$ be such an observable, and let the
system be in a state $\psi $ which, in order ``to be reached''
by a macroscopic observer, must be $\Gamma $-invariant (see the
preceding section).  Measuring an observable quantity
corresponding to $a$ when the system is in a $\Gamma $-invariant
state $\psi \in L^2(G_q)$ means to act with $a$ upon $\psi $.
The measurement will give as its result the eigenvalue $r_q$ as
determined by the eigenvalue equation
\begin{equation} \label{eigen1}
\pi_q (a)  \psi = r_q \psi 
\end{equation}
where, for simplicity we consider a non-degenerate case. Taking
into account the form of the representation $\pi_q$ (eq.
(\ref{Conrepr})) the above equation is equivalent to $$
\int_{G_q}a(\gamma_1)\psi (\gamma_1^{-1} \gamma ) = r_q \psi
(\gamma ).  $$ From the $\Gamma $-invariance of $\psi $ it
follows that  $\psi $ is constant on $G_q$; therefore, we can
write $$
\psi(\gamma_1^{-1} \gamma)\int_{G_q}a(\gamma_1) = r_q\psi
(\gamma ) $$ and consequently $$ r_q=\int_{G_q}a(\gamma_1).  $$
We have proved the following fact:
\par
{\em Lemma.\/} If $\psi \in L^2(G_q)$ is $\Gamma $-invariant and
if it is an eigenfunction of $a \in {\cal E}_H$, the eigenvalue
of $a$ is $r_q = \int_{G_q}a(\gamma_1)$.
\par
This is a nice conclusion. Let us notice that the result $r_q$
of a measurement is a measure in the mathematical sense (in this
case we deal with the Haar measure on the group $\Gamma $). But
we can go even further. Let us define the ``total phase space''
of our system $$ L^2(G) := \bigoplus_{q \in \Gzero }L^2(G_q), $$
with the operator $$
\pi (a) := (\pi_q(a))_{q \in G^{(0)}}
$$ acting on it. Now, eigenvalue equation (\ref{eigen1}) can be
naturally written as $$
\pi(a) \psi = r \psi
$$ where $r$ is a function on $G$; since , however, $\psi $ is
$\Gamma $-invariant $r$ can be interpreted as the function on
space-time $M$ $$ r: M \rightarrow {\bf R} $$ defined by $$ r(x)
= r_q = \int_{G_q}a(\gamma_1) $$ with $x$ being a point in $M$
to which the ``frame'' $q$ is attached. The measurement result
is not a ``naked number'', but a value of a function at a given
point $x \in M$ the domain of which is the entire space-time
$M$. As the consequence of this, if the measurement is performed
at a point $x \in M$, its result is correlated with the result
of another measurement performed on another component of the
same system even if it is situated at a very distant point $y
\in M$. This also suggests that typically quantum gravitational
phenomena should be looked for among correlations between
distant measurements rather than among ``local phenomena''. This
could be regarded as a relic of the pre-Planckian era in which
``everything was global''. It would be interesting to examine
nonlocal phenomena of quantum physics detected by the Aspect
type experiments in the light of the above remarks.
\par

\section{Concluding Remarks}
In the present paper we have proposed a mathematical structure
which combines essential aspects of general relativity (its main
geometric elements) with those of quantum physics (algebra of
operators of a Hilbert space), and leads to correct ``special
cases'' of standard general relativity and standard quantum
mechanics. In our opinion, this is the main result of our
approach. It is rooted in the fact that we have performed the
quantization of a groupoid (over a space-time) rather than of
space-time itself.
\par
Although we do not think that our approach is a full theory of
quantum gravity (it is a scheme rather than a theory since two
of its important elements, the group $\Gamma $ and the Fredholm
operator $F$, remain unspecified), it gives some hints of what
the future theory could be like. It clearly suggests that
measurable effects of quantum gravity should be looked for among
correlations of distant phenomena rather than among local
effects (Sec. 7). It is also a remarkable fact that a simple
model based on the groupoid $G = E \times D_4$ (Sc. 6) predicts
the value of ``cosmological constants'' which, after projecting
down to space-time, vanish.
\par
To choose the correct group $\Gamma $ it is important to
experiment with various possibilities. In \cite{HSL} we have
computed an example where $\Gamma $ is any finite group. A
generalization to the case when $\Gamma $ is compact would not
be difficult. From the physical point of view the most
interesting are cases with $\Gamma $ non-compact but then,
unfortunately, difficulties increase dramatically. We hope,
however, that they are only of the technical nature. For
instance, as the work by Fell \cite{Fell} demonstrates, the
group algebra of SL(2,C) (which physically is very interesting)
can be expressed in terms of algebras of operator fields on a
locally compact Hausdorff space, which in turn is isomorphic
with the algebra of norm-continuous functions on a certain
parameter space.
\par


\vspace{1cm}
\begin{thebibliography}{99}
\bibitem{Henry}
J. Demaret, M. Heller and D. Lambert, ``Local and Global
Properties of the World'', {\em Foundations of Science\/} {\bf
2}, 137-176 (1997).
\bibitem{ConRov}
A. Connes and C. R. Rovelli, ``Von Neumann Algebra Automorphisms
and Time-Thermodynamics Relation in Generally Covariant Quantum
Theories'', {\em Class. Quantum Grav.} {\bf 11}, 2899-2917
(1994).
\bibitem{Rovelli1}
C. Rovelli, ``Quantum Mechanics without Time: A Model'', {\em
Phys. Rev.} {\bf D42}, 2638-2646 (1990).
\bibitem{Rovelli2}
C. Rovelli, ``Time in Quantum Gravity: An Hypothesis'', {\em
Phys.  Rev.} {\bf D43}, 442-456 (1991).
\bibitem{ConLot}
A. Connes and J. Lott, "Particle Models and Noncommutative
Geometry", {\em Nucl. Phys. Suppl.\/} {\bf B18}, 29-47 (1990).
\bibitem{Cham}
A.H. Chamseddine, G. Felder and J. Fr\"ohlich, "Gravity in
Noncommutative Geometry", {\em Comun. Math. Phys.\/} {\bf 155},
205-217 (1993).
\bibitem{Sitarz}
A. Sitarz, "Gravity form Non-Commutative Geometry", {\em Class.
Quantum Grav.\/} {\bf 11}, 2127-2134 (1994).
\bibitem{Hajac}
P.M. Hajac, "The Einstein Action for Algebras of Matrix Valued
Functions - Toy Model", \JMP \ {\bf 37}. 45-49-4556 (1996).
\bibitem{ConG}
A Connes, "Gravity Coupled with Matter and the Foundation of
Non-Commutative Geometry", {\em Commun. Math. Phys.\/} {\bf 182},
155-176 (1996).
\bibitem{ChamCon1}
A.H. Chamseddine and A. Connes, {\em The Spectral Action
Principle\/}, Preprint hep-th/9606001 (1996).
\bibitem{ChamCon2}
A.H. Chamseddine and A. Connes, {\em A Universal Action
Formula\/}, Preprint hep-th/9606056 (1996).
\bibitem{MadMour}
fJ. Madore and J. Mourad, {\em Quantum Space-Time and Classical
Gravity\/}, Preprint LPTHE, Orsay 96/56 (1996).
\bibitem{Fuzzy}
J. Madore, {\em Gravity on Fuzzy Space-Time\/}, ESI Preprint 478
(1997).
\bibitem{HSL}
M. Heller, W. Sasin and D. Lambert, ``Groupoid Approach to
Noncommutative Quantization of Gravity'', \JMP \ {\bf 38},
5840-5853 (1997).
\bibitem{Connes}
A. Connes, {\em Noncommutative Geometry\/}, (Academic Press, New
York, 1994).
\bibitem{Madore}
J. Madore, {\em An Introduction to Noncommutative Differential
Geometry and Its Physical Applications\/} (Cambridge University
Press, 1955).
\bibitem{Masson}
T. Masson, {\em G\'eometrie non commutative et applications \`a
la th\'eorie des champs\/}, Thesis, preprint ESI 296 (Vienna,
1996).
\bibitem{HelSasSing}
M. Heller and W. Sasin, ``Noncommutative Structure of
Singularities in General Relativity'', \JMP \ {\bf 37},
5665-5671 (1996).
\bibitem{Renault}
J. Renault, {\em A Groupid Approach to $C^*$-Algebras\/}.
Lecture Notes in Mathematics, No 793 (Springer, Berlin, 1980).
\bibitem{Koszul}
J. L. Koszul, {\em Fibre bundles and differential geometry\/}
(Tata Institute of Fundamental Research, Bombay, 1960).
\bibitem{Dubois-Violette}
M. Dubois-Violette, {\em Some Aspects of Noncommutative
Differential Geometry\/}, preprint q-alg/9511027, 1995.
\bibitem{DubViol} M. Dubois-Violette, ``D\`erivations et calcul
diff\'erentiel non commutatif'', {\em C. R. Acad. Sci. Paris\/}
{\bf 307}, 403-408 (1988).
\bibitem{DubAl1}
M. Dubois-Violette, R. Kerner and J. Madore, ``Non-Commutative
Differential Geometry of Matrix Algebras'', \JMP \ {\bf 31},
316-322 (1990).
\bibitem{DubAl2}
M. Dubois-Violette, R. Kerner and J. Madore, ``Classical Bosons
in Noncommutative Geometry'', {\em Phys. Lett.} {\bf B217}, 485
(1989).
\bibitem{DubAl3}
M. Dubois-Violette, R. Kerner and J. Madore, ``Classical Bosons
in Noncommutative Geometry'', {\em Class. Quantum Grav.} {\bf
6}, 1709-1724 (1989).
\bibitem{DubMichor}
M. Dubois-Violette and P.W. Michor, ``D\`erivations et calcul
diff\'erentiel non commutatif, II'', {\em C. R. Acad. Sci.
Paris\/} {\bf 319}, S\'erie I, 927-931 (1994).
\bibitem{SH}
W. Sasin and M. Heller, ``Non-Commutative Differential
Geometry'', {\em Acta Cosmol. (Cracow)} {\bf 21}, no 2, 235-245
(1995).
\bibitem{Stephani}
H. Stephani, {\em General Relativity\/} (Cambridge University
Press, 1982).
\bibitem{Thirring}
W. Thirring, {\em Lehrbuch der Mathematischen Physik\/}, vol. 3
(Springer, Wien, 1979).
\bibitem{ConSur} A. Connes, ``Sur la th'eorie non commutative de
l'int'gration'', in: {\em Alg\`ebres d'Operateurs}, Lecture
Notes in Mathematics, no 725, (Springer, Berlin 1979) 19-143.
\bibitem{Palais}
R. S. Palais, {\em Real Algebraic Differential Topology\/}
(Publish or Perish, Wilmington, 1981).
\bibitem{HelSasStruct}
M. Heller and W. Sasin, ''Structured Spaces and Their
Application to Relativistic Physics'', \JMP \ {\bf 36},
3644-3662 (1995).
\bibitem{HelSasBanach}
M. Heller and W. Sasin, ``The Closed Friedman World Model with
the Initial and Final Singularities as a Non-Commutative
Space'', {\em Banach Center Publications\/} {\bf 41}, 153-162
(1997).
\bibitem{Serre}
P. Serre, {\em Repr\'esentations lin\'eaires des groupes
finis\/} (Hermann, Paris, 1978).
\bibitem{3Panie}
Y. Choquet-Bruchat, C. DeWitt-Morette and M. Dillard-Bleick,
{\em Analysis, Manifolds and Physics}, revised edition
(North-Holland, Amsterdam, 1982).
\bibitem{Fell}
J. M. G. Fell, ``The Structure of Algebras of Operator Fields'',
{\em Acta Mathematica}, {\bf 106}, 233-280 (1961).

\end{thebibliography}
\end{document}